\documentclass{JHEP3}

\usepackage{xspace}
\usepackage{multirow}
\usepackage{epsfig}
\graphicspath{{./figures/}}

\newcommand{\superk}      {Super-Kamiokande\xspace}       

\newcommand{\nue}         {$\nu_{e}$\xspace}
\newcommand{\numu}        {$\nu_{\mu}$\xspace}

\newcommand{\nusterile}   {$\nu_{sterile}$\xspace}
\newcommand{\mutau}       {$\nu_\mu \rightarrow \nu_{\tau}$\xspace}

\newcommand{\degree}      {$^\circ$\xspace}

\newcommand{\pizero}      {$\pi^{0} $}

\newcommand{\nuebar}      {$\overline{\nu}_{e}$\xspace}

\title{ The physics impact of proton track identification in future
  megaton-scale water Cherenkov detectors}

\author{M. Fechner\\Department of Physics, Duke
  University, Durham NC 27708, USA\\Present address: CEA, Irfu, SPP,
  Centre de Saclay, F-91191, Gif-sur-Yvette, France}
\author{C.W. Walter\\Department of Physics, Duke
  University, Durham NC 27708, USA}

\date{\today}

\abstract{
  In this paper, we investigate the impact in future megaton-scale
  water Cherenkov detectors of identifying proton Cherenkov rings.  We
  estimate the expected event rates for detected neutral current and
  charged current quasi-elastic neutrino interactions from atmospheric
  neutrinos in a megaton-scale Super-Kamiokande-like detector with
  both 40\% and 20\% photo-cathode coverage. With this sample we
  examine the prospects for measuring the neutrino oscillation
  pattern, and searching for sterile neutrinos.  We also determine the
  size of selected charged current quasi-elastic samples in a 300-kton
  fiducial volume Super-Kamiokande-like detector from examples of both
  conventional super-beams and beta-beams proposed in the literature.
  With these samples, it is shown that full kinematic neutrino
  reconstruction using the outgoing proton can improve the
  reconstructed energy resolution, and give good neutrino versus
  anti-neutrino tagging capabilities, adding important capabilities to
  water Cherenkov detectors in future projects. We determine the beam
  parameters necessary to make use of this technique and present
  distributions of neutrino and anti-neutrino selection efficiencies.
}

\begin{document}

\section{Introduction}

In~\cite{proton}, it was shown that the particle identification (PID)
algorithm of the \superk large water Cherenkov
detector~\cite{fukuda:2002uc} could be extended to identify
protons. This new tool was used to select single track proton events,
mostly produced in neutral current (NC) elastic collisions of atmospheric
neutrinos with protons in the water. Proton identification was also
used to tag charged-current quasi-elastic (CCQE) atmospheric neutrino
events.  The importance of being able to identify CCQE events in a
water Cherenkov detector is two-fold: it allows full kinematic
reconstruction of the neutrino track, and since CCQE proton production
occurs only for neutrinos and not anti-neutrinos, it selects a
quasi-pure neutrino sample.

Currently, many new ideas for neutrino experiments are being explored.
After the current generation of experiments, new projects to search
for CP violation in the neutrino sector and measure the neutrino mass
hierarchy may be
undertaken~\cite{issphysics,Abe:2007bi,Barger:2007yw}.  These
experiments utilize different beams, but all make use of very large
detectors.  In addition to the neutrinos that come from beams,
these large detectors can also study atmospheric and other
naturally-produced sources of neutrinos.  The capabilities of each
detector technology, and the requirements needed to utilize them fully
are key issues in designing new facilities.  The
identification of proton rings is a new capability in water Cherenkov
detectors which should be considered when designing these experiments.

Several of the implications of having proton identification in a water
Cherenkov detector have been previously explored
in~\cite{Beacom:2003prd}. In this work, we use the \superk simulation
and reconstruction algorithms to make detailed predictions of the
event rate and performance of this technique in future large water
Cherenkov detectors exposed to atmospheric and intense artificial
beams. We also provide information on efficiencies of detection and
photo-cathode coverage requirements which should be of use in future
simulation studies.

\section{Summary of the proton identification technique}
\label{sec:tech}
In this work we only briefly summarize the main points of the
identification of Cherenkov rings from protons. For a complete
description the reader should consult~\cite{proton}. 

The main difficulty in identifying protons is separating them from
muon tracks, which can have similar ring
characteristics, in particular sharp ring edges. Protons tend to
interact early in the water, producing shorter tracks than muons. Due
to their heavier masses, they also produce Cherenkov rings with
smaller opening angles than muons. These observations are used in the
separation method.  Proton identification is a hypothesis test, used
to accept or reject the hypothesis that the observed Cherenkov ring
pattern on the photomultiplier tubes (PMTs) of the detector was caused
by a proton. A ratio of maximum likelihoods is built from the
likelihood of the particle to be a proton, and then a muon. It is then
used along with other discriminating variables to make a decision
about the particle type.

The first stage of all event processing is a fit of the event vertex,
followed by a search for all visible Cherenkov rings. For \superk
these steps are described in e.g.~\cite{ashie:2005ik}.  At this stage,
the vertex, track direction and Cherenkov cone's opening angle are
known, allowing particle identification to be attempted. In the case
of a single ring event, the fitted track parameters and the trial particle
types are used as inputs to compute the mean expected charges collected by each
PMT. This produces the ``expected light pattern'' of the track
configuration under study. This calculation relies on Cherenkov light
density tables, which were pre-computed for the relevant particle
types using intensive Monte-Carlo simulations. The likelihood of the
observed pattern to an expected light pattern can be calculated, and
is referred to as a ``pattern likelihood''.  For the proton
hypothesis, the pattern likelihood is maximized by adjusting the
momentum and track length, yielding the maximum pattern likelihood 
$\mathcal{L}_p$. Then, for the muon hypothesis, a fit of
the observed light pattern to a muon's expected light pattern is
performed, yielding the maximum pattern likelihood $\mathcal{L}_\mu$.  
The hypothesis test relies on the ratio of the maximum
likelihoods $\mathcal{L}_p\over\mathcal{L}_\mu$, as well as the 
best fit estimates of the proton momentum
and track length \cite{proton}.

For CCQE event identification, this method is extended to handle
two-prong events, with one lepton and one proton. In principle these
should be two-ring events. However studies have shown that CCQE events
can confuse the ring finding algorithm: because of the weakness of
proton rings when the particle is just above threshold, the
reconstruction algorithm often only finds the lepton, although
eye-scanning reveals a clear second ring.  Therefore CCQE events can
be reconstructed either as two-ring or single-ring events.  When the ring
finding algorithm has identified two rings the hypothesis that the
second identified ring is a proton is tested by applying a similar
method to the one described above, superimposing lepton and proton
light patterns. For events reconstructed as single ring, the situation
is more complicated, since the missing ring must first be
identified. A dedicated ring fitter incorporating the proton
identification technique was developed and applied to single events,
thereby doubling the tagged-CCQE sample size in the atmospheric neutrino
sample.

Finally, a set of stringent selection cuts is applied to reduce
non-proton backgrounds in the CCQE sample.  In order to reduce the
non-CCQE background, the full kinematics of the reconstructed final
state is used. We use the variable $V^2=(p_p+p_l-p_N)^2$, where $p_p$,
$p_l$ and $p_N$ are (resp.) the four-momenta of the proton, lepton,
and target neutron (assumed to be immobile).  Assuming that the
reaction is indeed CCQE, a selection cut on $V^2$ of the outgoing
proton-lepton system is made.  For a true CCQE event, $V^2$ is the
invariant mass of the incoming neutrino and should then be close to
zero.  Under the CCQE assumption, non-CCQE events with misidentified
or missing particles will have a non-zero invariant mass. Further
details on the selection cuts can be found in \cite{proton}.

This technique was verified by applying it to the \superk atmospheric
neutrino data set corresponding to 141 kton\,years of exposure.  Both
single protons and CCQE event samples were selected and good agreement
was found between data and Monte Carlo. Additionally, a fit to the
atmospheric oscillation parameters using this data set found agreement
with previously published results~\cite{proton}.

In the remainder of this paper, we will call ``tagged-CCQE'' events
those events with a lepton-like ring and a proton-like ring that pass
the selection cuts outlined above. 

\section{Observability conditions of protons in a water Cherenkov
  Detector}
\label{sec:visib}

Very close to the Cherenkov threshold, protons can make very weak
rings that cannot be detected well.  In order to be above Cherenkov
threshold, and make enough light to be visible in the detector, the
proton momentum must be at least 1.1 GeV/c~\cite{proton}.  The minimum
neutrino energy required to produce such a proton from a CCQE
collision in water is approximately 1 GeV. Thus, one requirement for
physics studies using CCQE tagging is a high enough event rate above 1
GeV.  The efficiency of detection depends on many parameters,
including distance from the vertex to the walls of the detector, and
is therefore not trivial to estimate. Using a full simulation which
accounted for all of these effects, we have produced the histogram with
triangular markers shown in figure~\ref{fig:eff}.  The rising response
shows the ``visibility'' of mono-energetic protons in the \superk
detector. As expected, it turns-on sharply around 1100 MeV/c and
increases with momentum to reach almost 100\%.

However, there is a competing effect: above proton momenta of $\approx
2$ GeV/c, which corresponds to neutrino energies above a few GeV,
protons often produce secondary particles through hadronic
interactions with the water. These secondaries (charged pions or
showers from neutral pions) emit Cherenkov light themselves, and impede
proper reconstruction due to the presence of extra Cherenkov rings in
the event. This probability of producing visible secondaries in the
water increases with proton momentum, and makes it almost impossible
to identify protons above $\approx 2.5$ GeV/c. In
figure~\ref{fig:eff}, we also show the fraction of protons that do not
produce any visible secondary as a function of momentum (the histogram
with square markers) calculated using mono-energetic proton
Monte-Carlo events.

Taken together, these two constraints restrict the bounds over which the
present technique works well to within the $1.2-2.0$ GeV/c
momentum range.  Therefore, it is not applicable to all beam spectra;
it only works for incoming neutrinos with energies of a few GeV.

Also shown in figure~\ref{fig:eff} is the efficiency of the
reconstruction technique for CCQE events (crosses), which we
calculated using atmospheric neutrino Monte-Carlo events.  As expected
from the previous two effects, it peaks around 1.3 GeV/c and remains
appreciable until $\approx 1.9 $ GeV/c.  The efficiency of the CCQE
tagging method shown in figure~\ref{fig:eff} falls to zero more
quickly than the combined effect of ``visibility'' and hadronic
interactions because proton patterns become very similar to other
track patterns at higher momenta. They therefore fail to pass the
selection cuts based on the likelihood fits described above, which
further reduces the efficiency even in the absence of secondaries.

\FIGURE{\epsfig{file=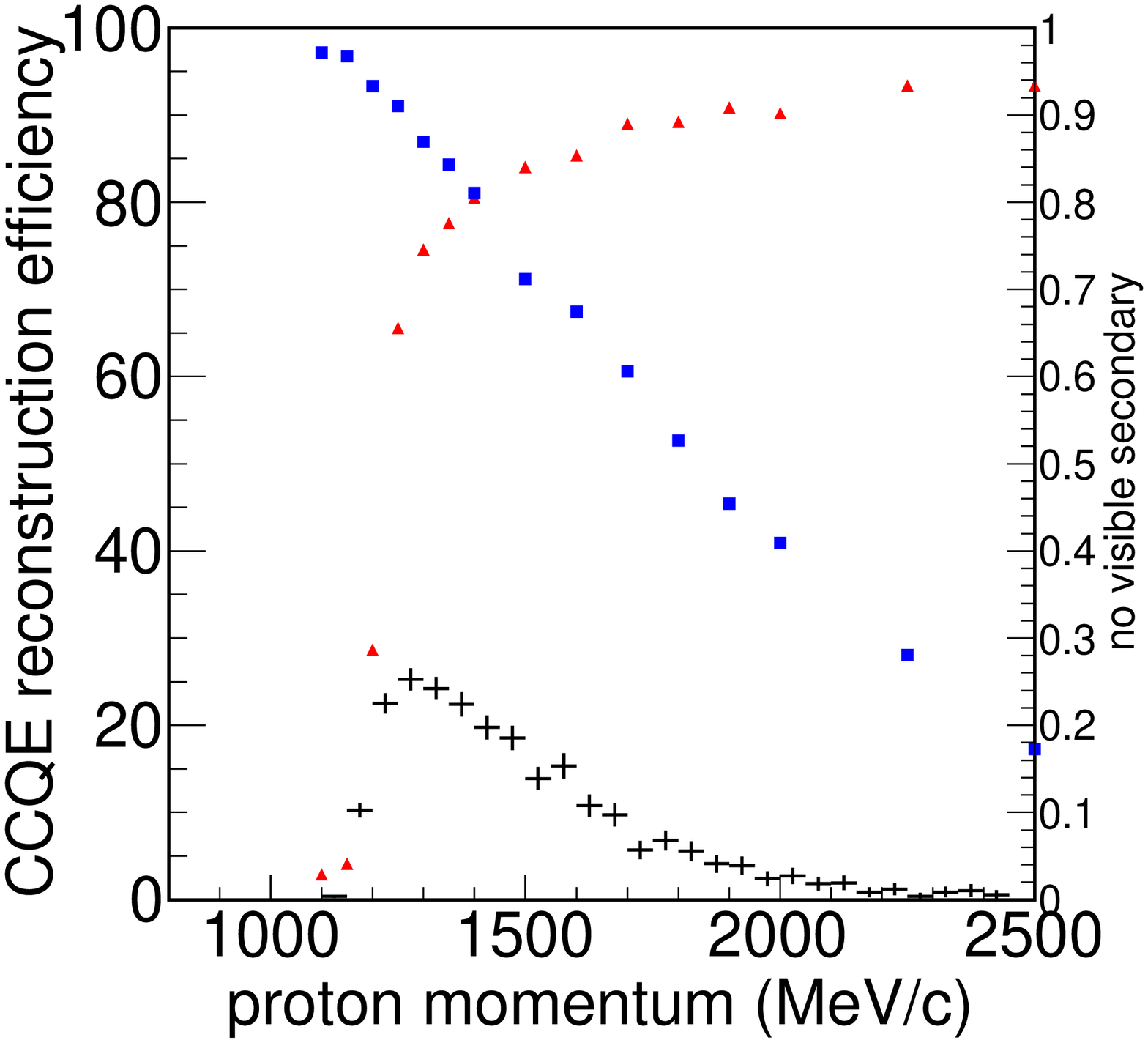,width=3.5in}
  \caption{Efficiency of CCQE reconstruction as a function of true
    proton momentum in MeV/c. The triangles show the fraction of
    single protons that are visible as a function of proton momentum.
    This curve has a sharp turn on at the Cherenkov threshold. The
    square marks show the fraction of protons that do not create
    visible secondaries in the water, which decreases with momentum,
    and should be read on the axis on the right hand-side. Those
    two graphs were calculated using Monte-Carlo simulations of single
    protons in the detector. The graph with cross markers
    shows the detection efficiency for
    CCQE event identification as a function of proton momentum,
    obtained with the full simulation and reconstruction of simulated
    atmospheric neutrino events. At higher momenta it is lower than 
    the combined effect of visibility and secondary production
    because identifying the ring pattern as a proton becomes very
    difficult as momentum increases.\label{fig:eff}}}

\section{Neutrino versus anti-neutrino tagging by proton identification}
\label{sec:nunubar}

Another great benefit of CCQE tagging by proton identification is
that it provides a means of selecting an almost pure neutrino
sample. This is due to the fact that in CCQE collisions only neutrinos
produce protons in the final state; anti-neutrinos will produce
neutrons.  The presence of a proton does not guarantee the nature of
the incoming neutrino, since other anti-neutrino interaction channels
as well as hadronic interactions of neutrons can produce protons. But,
even in non-QE events the method preferentially selects neutrino
interactions because it picks up protons (and not neutrons) in the
final state.  Monte-Carlo studies from \cite{proton} show that for the
\superk atmospheric $\nu$ data set, the neutrino fraction of the
sample after CCQE selection cuts is $91.7\pm3(\mathrm{syst})\%$.
Therefore the tagged-CCQE sample is an almost pure neutrino sample
when used with atmospheric neutrinos.

Using our Monte-Carlo samples we have calculated the $\nu$ and
$\bar{\nu}$ selection efficiencies as a function of neutrino energy,
i.e.~the ratio of all events selected in the tagged-CCQE sample to all
events that occur in the detector's fiducial volume (both CCQE and
non-CCQE interaction modes are included).  These efficiencies are
shown in figures~\ref{fig:sk1eff40} and \ref{fig:sk1eff20}, and can 
be used as input for further
simulations. We show the efficiencies for two different values
of the photo-cathode coverage of the detector, as this parameter is 
important for future detector design (see section~\ref{sec:atmlarge}).
The peak fraction is $\approx 2\%$ for neutrinos and
$\approx 0.5\%$ for anti-neutrinos. It is quite low because all
detection effects relevant to water Cherenkov detectors are included
(data reduction, vertex fitting, ring counting and finally proton
selection cuts), along with the high Cherenkov threshold of
protons. However, as will be seen, with large exposures, tagged
samples of hundreds of events can be expected in future facilities.

\FIGURE{\epsfig{file=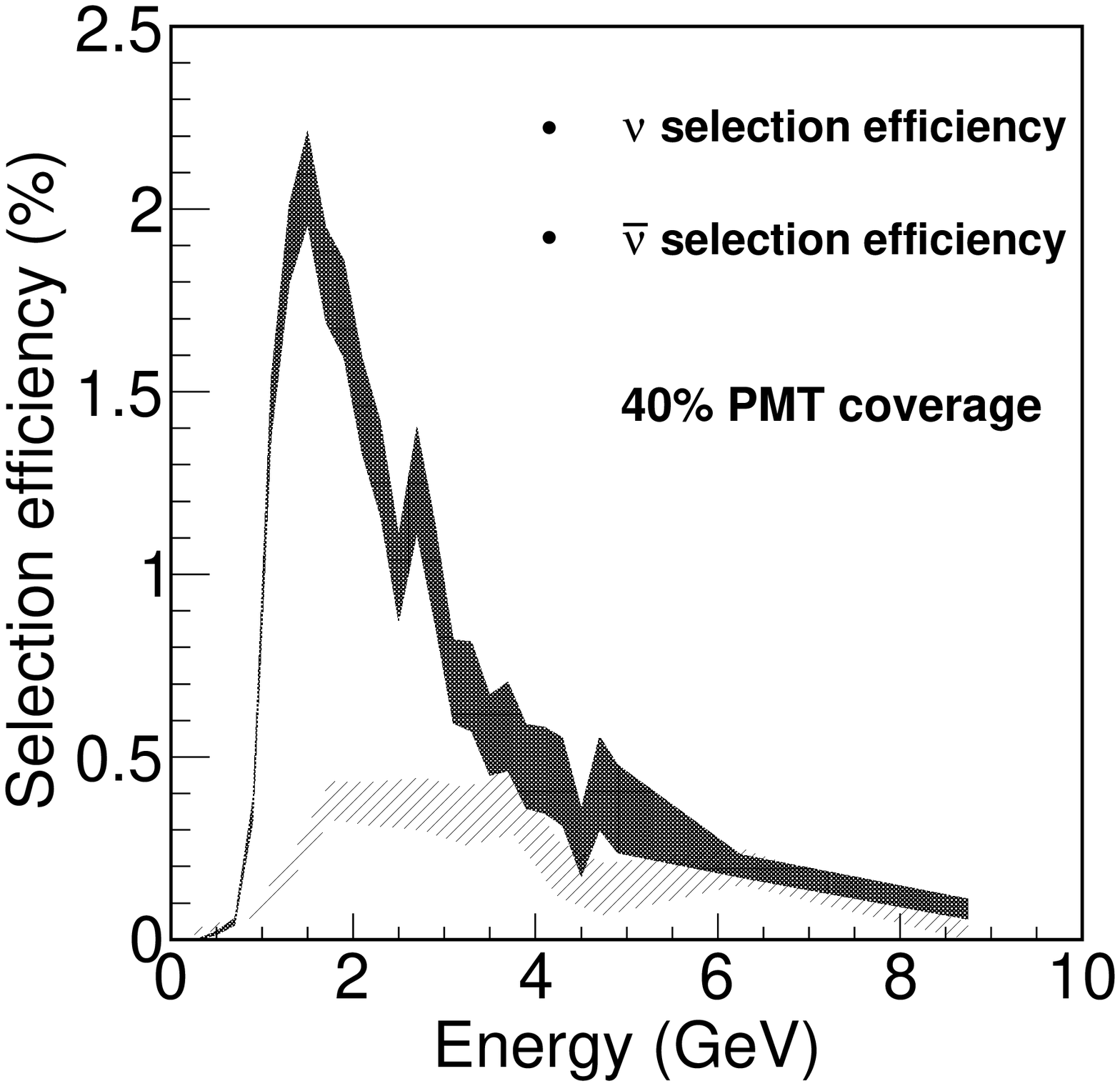,width=4in}
  \caption{Neutrino and anti-neutrino selection efficiencies as a
    function of the incoming neutrino energy for 40\% photo-cathode
    coverage (see section~\ref{sec:atmlarge} for discussion of photo-cathode coverage).\label{fig:sk1eff40}}
}

This capacity to select neutrinos as opposed to anti-neutrinos will
potentially be useful for studies of CP odd matter effects, 
especially with regard to the neutrino mass hierarchy.
%if sufficiently large due to matter effects. 
Previous studies have investigated the feasibility of such
approaches with magnetized iron calorimeters, to separate neutrinos and
anti-neutrinos~\cite{Indumathi:2004kd}.  
The new quasi-pure neutrino samples provided by proton tagging open up
similar possibilities for large water Cherenkov detectors, although
they allow only the selection of neutrino, not pure anti-neutrino
samples.

\FIGURE{\epsfig{file=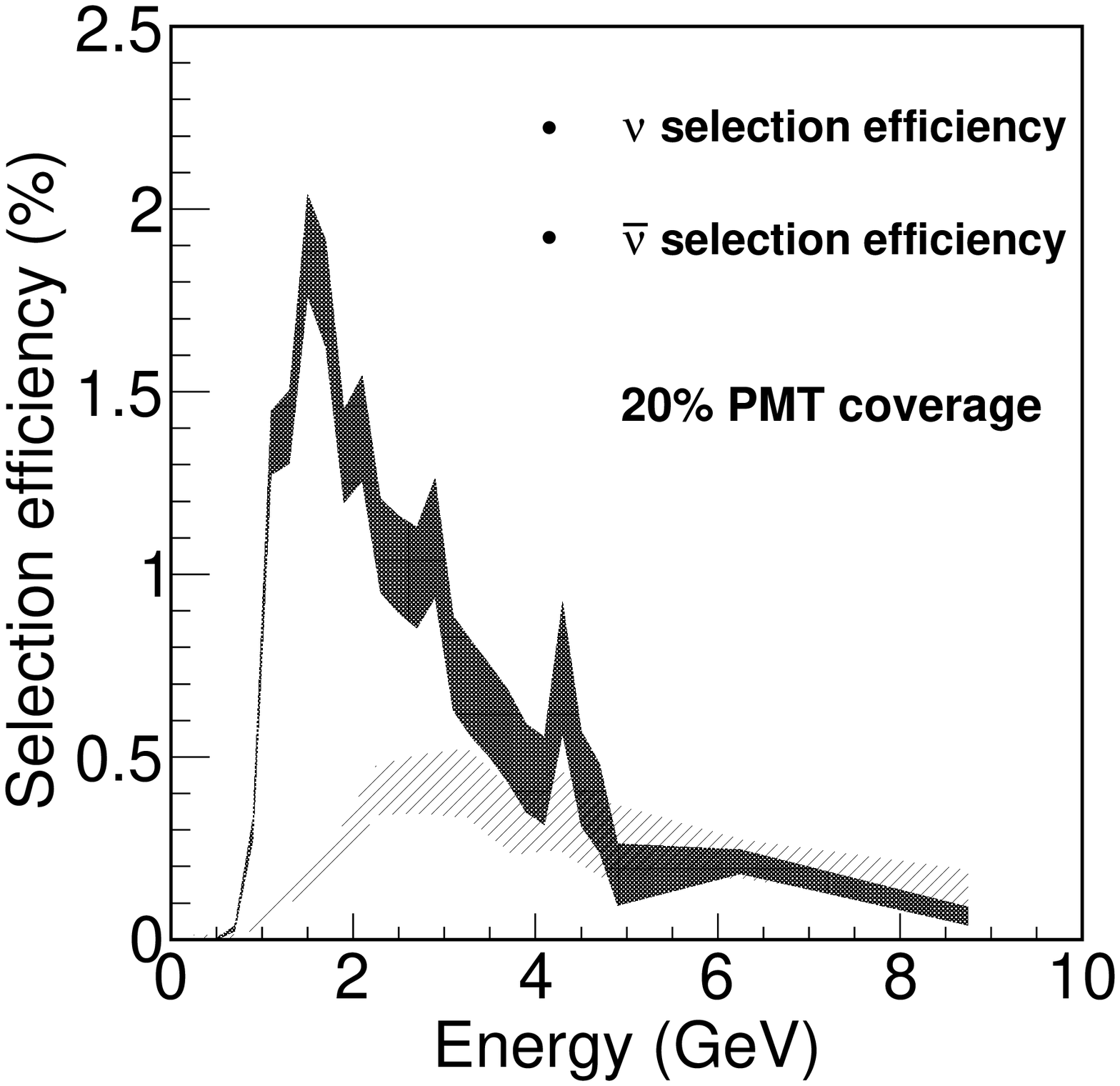,width=4in}
  \caption{Neutrino and anti-neutrino selection efficiencies as a
    function of the incoming neutrino energy for 20\% photo-cathode coverage (see section~\ref{sec:atmlarge} for discussion of photo-cathode coverage).\label{fig:sk1eff20}}
}

\section{Future large water Cherenkov detectors}
\label{sec:atmlarge}

Several possible future megaton-scale water Cherenkov detectors have
been, or are currently being studied in different countries, see for
example~\cite{memphys,uno,hyperk,Diwan:2006qf,Barger:2007yw}.  For a
recent review and comparison of many of these options the reader is
referred to~\cite{Abe:2007bi}.  Using the information gained from our
simulations with the \superk detector we have made predictions on the
rate of fully reconstructed CCQE events in a water Cherenkov detector
for both atmospheric neutrinos and several future intense neutrino
beam configurations. We also studied the use of this technique
to tag neutrino (as opposed to anti-neutrino) events.

An important question to address when designing a large water
Cherenkov detector is the photo-cathode coverage, namely the fraction
of the total area of the detector's wall which is instrumented for
Cherenkov light collection. Higher coverages induce higher costs. For
this purpose, in the studies presented below, we have used the SK
simulation to simulate both 40\% and 20\% photo-cathode coverage.  The
higher coverage uses the run configuration of 1996-2001, known as
SK-I. The lower coverage uses the simulation setup from the run period
of 2003 to 2005 (SK-II), following partial reconstruction after an
accident in November 2001 that destroyed half the tubes.  In what
follows we assume that the simulated detectors have the same response as
Super-K with these two configurations.

\section{Atmospheric neutrinos}

\subsection{Event rates of proton-tagged atmospheric neutrinos}

For determining the expected event rates and detector response from
the atmospheric neutrino flux in a future large water Cherenkov
detector, the simulations of the \superk detector can simply be scaled
by exposure.  For an exposure of 1 Mton\,yr, the gain in statistics
would be a factor of $\approx 7$ compared to the \superk data set
(141~Kton\,yr). Table~\ref{tab:summary} summarizes projected estimates
of event rates for 1 Mton\,yr of exposure to atmospheric neutrinos.
The predictions for both single-ring proton-like events and 
tagged-CCQE events are given.  

\TABLE
  {\begin{tabular}{lcc}
    \hline
    \hline
    \multirow{2}{100pt}{Event class}  & Expected in 1 Mton\,yr & Expected in 1 Mton\,yr\\
    &  (40\% coverage)  &  (20\% coverage) \\
    Single proton   & 375 & 310 \\
    Tagged CCQE e-like     & 337 (53.0\%) & 295 (51.4\%) \\
    Tagged CCQE $\mu$-like & 500 (62.4\%) & 450 (61.3\%) \\
    \hline 
    \hline
  \end{tabular}
  \caption{Summary of the projected data samples for atmospheric
    neutrinos: Single proton, tagged-CCQE e-like and tagged-CCQE
    $\mu$-like, for a Mton-scale detector with efficiencies similar to
    SK. For tagged-CCQE events the number in parentheses is the
    fraction of true CCQE events in the selected samples estimated
    from our Monte-Carlo simulation.}
  \label{tab:summary}
}

It can be seen in table~\ref{tab:summary} that a reduction from 40\% to 20\%
photo-cathode coverage would lead to an overall event loss of about
10\% for CCQE searches, and about 20\% for NC elastic searches.

\subsection{Kinematic reconstruction of atmospheric neutrinos}

As shown in \cite{proton}, one of the main appeals of the CCQE tagging
technique is accurate kinematic reconstruction of the incoming
neutrino track. It is useful for $L/E$ reconstruction, where $L$ is
the neutrino flight path and $E$ its energy.  An important feature of
$L/E$ reconstruction is its potential for seeing the oscillation
shape, and therefore discriminating between oscillation models and
other models that could also explain a zenith-dependent flux
suppression. In~\cite{ashie:2004mr}, the \superk collaboration
performed an $L/E$ analysis using only the lepton momentum (no proton
tagging was available). That analysis saw evidence of the oscillatory
pattern with high significance.  In \cite{proton}, using the tagged
CCQE sample, the \superk collaboration performed another $L/E$
analysis using the reconstructed neutrino information, and found
consistent results, albeit with much lower statistics.

A kinematically reconstructed CCQE sample has good resolution in both
neutrino energy (approximately $15\%$) and direction (approximately 
12\degree for \numu and 16\degree for \nue) according to the studies shown
in \cite{proton}(see table IX).  One might naively think that
with the improved energy resolution of the proton-tagged CCQE sample
with high statistics it would be possible to see the sinusoidal
oscillation pattern with even higher resolution than observed
in~\cite{ashie:2004mr}. However, near the horizon,
$dL/d\cos\theta_{\mathrm{zenith}}$ is very large, and $L/E$ cannot be
precisely determined, even with this sample.
Figure~\ref{fig:le-resolution} shows where the first and second
oscillation maxima fall in $(\cos\theta_{\mathrm{zenith}}, E)$ (dotted
and dashed lines), as well as the region where the $L/E$ resolution is
worse than 70\% (using the same criterion as in~\cite{ashie:2004mr}) 
despite kinematic reconstruction (between the quasi-vertical
lines).  It can be seen that in the energy range spanned by our CCQE
atmospheric sample, as constrained by conditions explained in
section~\ref{sec:visib}, the first maximum occurs very near the
horizon, in a region where $L/E$ precision is low.  Even with a
Mton-scale Cherenkov detector, the first maximum would remain out of
reach. However the second maximum might be visible, but the statistics
would remain relatively low (69 expected $\mu$-like events with $-0.6
<\cos\theta_{zenith}<-0.2$ in 1 Mton\,yr).

\FIGURE{\epsfig{file=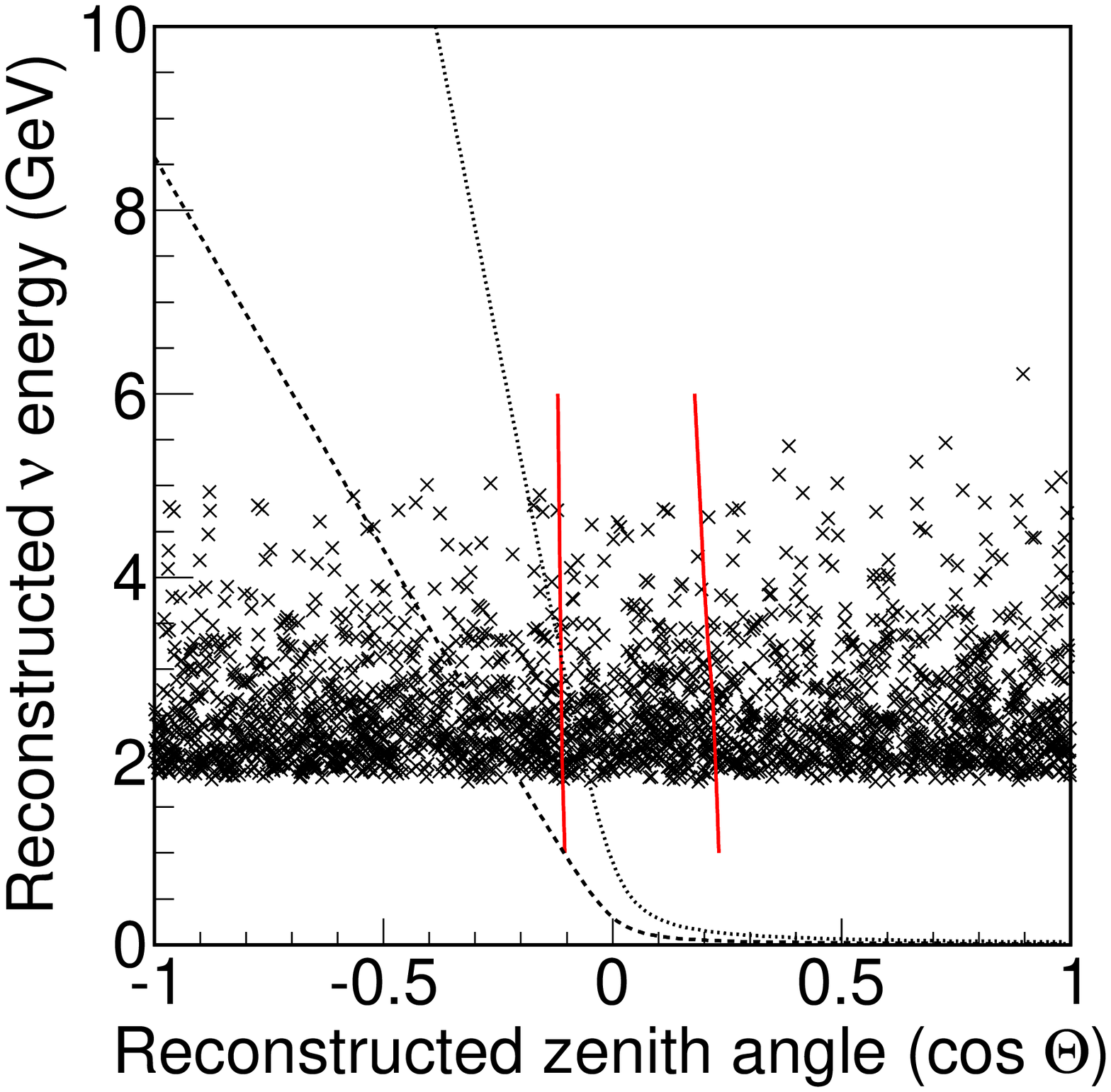,width=4in}
\caption{Study of the $L/E$ resolution as a function of reconstructed
  energy and zenith angle. The region inside the quasi-vertical lines
  corresponds to an $L/E$ resolution worse than 70\%. The dotted and
  dashed lines indicate the position of the first and second maximum
  of the oscillation (resp.). The scattered dots show the density of
  tagged-CCQE Monte-Carlo events at \superk.}
\label{fig:le-resolution}
}

\subsection{Search for sterile neutrinos using single-proton events}

In \cite{proton}, a sample of single ring proton events was selected,
and was shown to be comprised of $\approx 38\%$ of neutral current
elastic events $\nu+p\rightarrow\nu+p$. This sample has potential
sensitivity to sterile neutrinos, because the proton and the incident
neutrino directions are correlated (with a mean scattering angle of 
$\approx 40^\circ$). Muon neutrinos that are up-going travel a longer 
path through the Earth and oscillate mainly to $\nu_\tau$. Since
the neutral current cross-section is flavor-insensitive, the single
proton sample is mainly insensitive to \mutau oscillations. However,
if sterile neutrinos exist, and active neutrinos oscillate to sterile
neutrinos, a deficit of up-going single-proton events would be observed.
To test this hypothesis, the up-down asymmetry $\frac{U-D}{U+D}$
can be calculated. As in \cite{proton}, up-going events have
$-\cos\theta_{\mathrm{zenith}}<-0.2$ and down-going events 
have $-\cos\theta_{\mathrm{zenith}}>0.2$, 
where $\theta_{\mathrm{zenith}}$ is the angle between the proton
track and the vertical.

Based on Monte Carlo, the maximum asymmetry is $\approx
-15\%$~\cite{proton} and would occur if all neutrinos oscillated to
sterile neutrinos which has been previously ruled out
by~\cite{fukuda:2000np,Adamson:2008jh}.  However, models with sterile neutrino
admixtures are still the subject of investigation. In such models,
some of the \numu can oscillate into \nusterile in addition to the
active flavors of neutrinos.  With a 1 Mt\,year exposure, the sample
would contain approximately 150 up-going and down-going events (in the
absence of sterile neutrinos). The statistical uncertainty on the
asymmetry would still be on the order of 6\%. \superk has previously
studied the so-called 2+2 model (see \cite{Fogli:2000ir}), for which
an asymmetry of $\approx -6\%$ is reached for sterile admixtures of
$\approx 40\%$.  Using other samples, \superk has already ruled out
admixtures larger than 26\% at 90\% CL \cite{Walter:2003zj}, showing
that the single proton event sample does not have enough power on its
own to further constrain searches for sterile oscillations, even for a
1Mt\,year exposure.  However, this sample would be a helpful addition
to a combined search for sterile neutrinos which used the several
different samples sensitive to sterile neutrinos available in water
Cherenkov detectors.  The numbers quoted above were obtained at 40\%
photo-cathode coverage.  As shown above, at 20\% coverage the rates
are expected to be roughly 20\% lower.

\subsection{Summary of expectations with atmospheric neutrinos}

In summary, a very large water Cherenkov detector using proton
identification would observe several hundreds of NC elastic and CCQE
events.  The main additions to physics would be the potential
visibility of the second oscillation maximum in an $L/E$ analysis, as
well as the very high neutrino (as opposed to anti-neutrino) 
content of the selected CCQE sample, allowing
further studies of mass hierarchy effects with atmospheric neutrinos.
Sterile neutrino searches based exclusively on this sample would
likely not further constrain models beyond the present limits but
would be a valuable addition to more comprehensive searches.

\section{Artificially produced neutrino beams}
\label{sec:future}

Many new ideas for future neutrino experiments involving large
detectors have been studied extensively over the past few years, with
a variety of $\nu$ production and detection scenarios, including wider
band beams with longer baselines (see \cite{issphysics} and references
therein). In this section, we study the expected benefits of the
CCQE reconstruction method in three examples of upcoming or planned
neutrino beams. Our first example will be the T2K experiment's beam
which is located in Japan. Then, we study another example of
a conventional neutrino beam that could be produced at the future
Project-X~\cite{Goldenbook,ProjectX} accelerator complex at
Fermilab. Finally, we examine two possible beta-beam
configurations. The beam characteristics that we considered and the
expected performances are detailed below.

\subsection{Tokai to Kamioka (T2K)}

This experiment \cite{Itow:2001ee} will use a 2.5\degree off-axis
$\nu_\mu$ beam produced at Tokai (Japan) and detected at \superk
(fully rebuilt, with 40\% photo-cathode coverage) 295~km away. It will
begin in 2009.  However the off-axis angle is tuned for a spectrum
peak at 0.6 GeV, with the vast majority of neutrinos below 1 GeV, thus
making the number of visible protons very small. We estimate that only
24.4 CCQE events would have a visible proton for $5\times 10^{21}$
protons on target (40 GeV beam protons). With the same analysis as
in \cite{proton}, only 13.0 events would be selected, half of which would truly
be CCQE. Table~\ref{tab:t2k} shows the breakdown as a function of
lepton type. 

\TABLE{
  \begin{tabular}{lcc}
    \hline\hline
    lepton flavor & ~Selected~  & ~true-CCQE~   \\
    e-like        &      $2\pm 0.2$      & $0.5\pm 0.04$      \\
    $\mu$-like    &      $10.9\pm 0.3$   & $5.9\pm 0.2$      \\
    \hline\hline
  \end{tabular}
  \caption{Expected event selection in the T2K beam for $5\times
    10^{21}$ protons-on-target, after the selection criteria defined
    in \cite{proton}. The columns labeled true-CCQE show the expected 
    numbers of true CCQE events in the selected sample.
    We have assumed standard 3-flavor neutrino
    oscillations, with $\theta_{13}=0$.}
  \label{tab:t2k}
}

Therefore, we conclude that T2K cannot benefit from CCQE tagging
due to the very low statistics. More generally, for a narrow band beam
to make use of the CCQE selection technique presented here, protons
must be observable, thus the peak energy must be above $\approx 1$
GeV, but below a few GeV as explained above.

\subsection{The Fermilab to DUSEL Long Baseline Neutrino Experiment (LBNE)}
\label{sec:wbsb}

One of the future projects currently under study is a $\nu_\mu$ beam
produced at Fermilab (Illinois, USA), and aimed at the Homestake
(South-Dakota, USA) mine 1300 km away, the chosen site for
the future Deep Underground Science and Engineering Laboratory
(DUSEL). Generically, this experiment is known as the Long Baseline
Neutrino Experiment (LBNE).  If built, the DUSEL facility could host a
Mton-scale water Cherenkov detector. At Fermilab, planning is under
way to design a high intensity 2.3~MW neutrino beam based on the
Project-X high intensity proton source which could be directed to the
DUSEL laboratory. Project-X is built around a 8~GeV superconducting linac
which would be paired with modified versions of the existing
accelerator complex at Fermilab to make high intensity neutrino, kaon
and muon beams~\cite{Goldenbook,ProjectX}.

In the remainder of this section, we have assumed a project-X based
neutrino beam with a power of 2.3~MW, using 120~GeV protons. The beam
is a wide band beam, and the detector is 1300~km away from the source
and 12~km off-axis.  We also assumed that $3.6\times 10^{21}$ protons
were collected, which would correspond to 3 years of running with this 
beam at full power (about $3\times 10^7$ seconds of live time). 
The detector used in
these calculations is a water Cherenkov detector with a 300-kton
fiducial volume, which is assumed to have the same properties (event
reconstruction, efficiencies, systematics) as \superk.  The beam
fluxes are those of~\cite{Barger:2007yw, bishai}. 
The total integrated neutrino flux over the whole energy range (up to 120 GeV)
and 3-year running period is $4.5\times 10^9\ \mathrm{m}^{-2}$ for $\nu_e$ and $4.6\times 10^{11}\ \mathrm{m}^{-2}$ for $\nu_\mu$.
For the remainder
of this article we will refer to this configuration as the LBNE beam
or LBNE project when referring to the beam and detector together.

\subsubsection{Event rates and purity of CCQE tagged events}

Re-weighting the \superk atmospheric Monte-Carlo to the LBNE beam
spectrum, we estimate that 750 true-CCQE events with only a single
fitted ring will be produced, along with 500 two-ring true-CCQE events
(accounting for neutrino oscillations).  Using the same selection cuts
as for the atmospheric analysis described above, the total tagged-CCQE
sample (e-like and $\mu$-like) will contain 750 to 800 events
(depending on the PMT coverage), of which about 450 to 470 are truly
CCQE. We expect to select roughly 650 $\mu$-like events and 100 e-like
events in the tagged-CCQE sample.
%Table~\ref{tab:projectX} shows the number of events as a
%function of tagged lepton flavor, e-like or $\mu$-like. 
Although the spectrum extends to over 50 GeV, the neutrino energy is
below 5 GeV in the CCQE sample because of the visibility conditions
explained in section \ref{sec:visib}.
The average resolution on the measured neutrino energy for this sample
is $\approx 15\%$.

%\TABLE{
%  \begin{tabular}{lcccc}
%    \hline\hline
%    lepton flavor & ~Selected~ & ~true-CCQE~  & ~Selected~ & ~true-CCQE~ \\
%    photo-coverage & 40\% & 40\% & 20\% & 20\% \\
%    e-like       & $91\pm 15$   & $13 \pm 2$  &  $106 \pm 15$  & $20 \pm 4 $ \\
%    $\mu$-like   & $698\pm26$  & $464\pm 20 $  &   $648 \pm 26$ & $422 \pm 20 $ \\
%    \hline\hline
%  \end{tabular}
%  \caption{Expected event selection in the LBNE project after
%    the selection criteria defined in \cite{proton}. We have
%    assumed standard 3-flavor neutrino oscillations, with
%    $\theta_{13}=0$. 
%    The columns labeled true-CCQE show the expected 
%    numbers of true CCQE events in the selected sample.
%    The error bars correspond to Monte-Carlo
%    statistical error, and are large because of our re-weighting method.}
%  \label{tab:projectX}
%}

%As also seen in table~\ref{tab:summary} in section~\ref{sec:atmlarge},
In table~\ref{tab:summary} in section~\ref{sec:atmlarge}, we observed that
the selected event rate is reduced by approximately
10\% if 20\% photo-coverage is used while achieving almost the same
CCQE purity as the 40\% case. For the LBNE beam, the trend is consistent
but precise numbers are difficult to present owing to errors caused by
our re-weighting method.
%This is best seen in the $\mu$-like
%selection where the statistics are good enough to evaluate the
%efficiency.

High energy neutrino interactions are a background, especially for
\nue events: the low fraction of CCQE events in the e-like sample
is due to the large amount of neutral current \pizero\ production
induced by high energy neutrinos in the wide band beam; gamma showers
from \pizero\ decays fake CCQE \nue events. Extra techniques for
neutral pion background rejection could be applied to select a cleaner
\nue sample, but statistics would be even lower.  The value of
$\theta_{13}$ has a relatively modest influence on the number of
selected events: the variation of the e-like sample is on the order of
15\% when $\theta_{13}$ varies from 0 to $4^\circ$, near the expected
sensitivity limits of the current generation of experiments.

\subsubsection{Energy resolution of CCQE tagged events}

One important benefit of CCQE selection is direct neutrino energy
reconstruction, with a resolution of about 15\% for such a beam. 
Figure~\ref{fig:pxspec} shows the reconstructed
spectra for $\mu$-like events in the tank, with and without
oscillation.  Note that the technique used to kinematically
reconstruct the neutrino is different from that which has been used in
K2K~\cite{Ahn:2006zza}: here no knowledge of the beam direction is
needed since both outgoing particles are known. In
section~\ref{sec:tech} we mentioned that the parameter $V^2$
(invariant mass of the outgoing lepton system, subtracting the neutron
mass) was used to reduce non-CCQE background.  With a beam, this cut
could also be supplemented by a comparison between the reconstructed
neutrino direction and the incoming beam direction to potentially
improve non-CCQE rejection. A full study with this requirement would
require a dedicated beam flux simulation rather than re-weighted
atmospheric Monte-Carlo.

\FIGURE{
  \epsfig{file=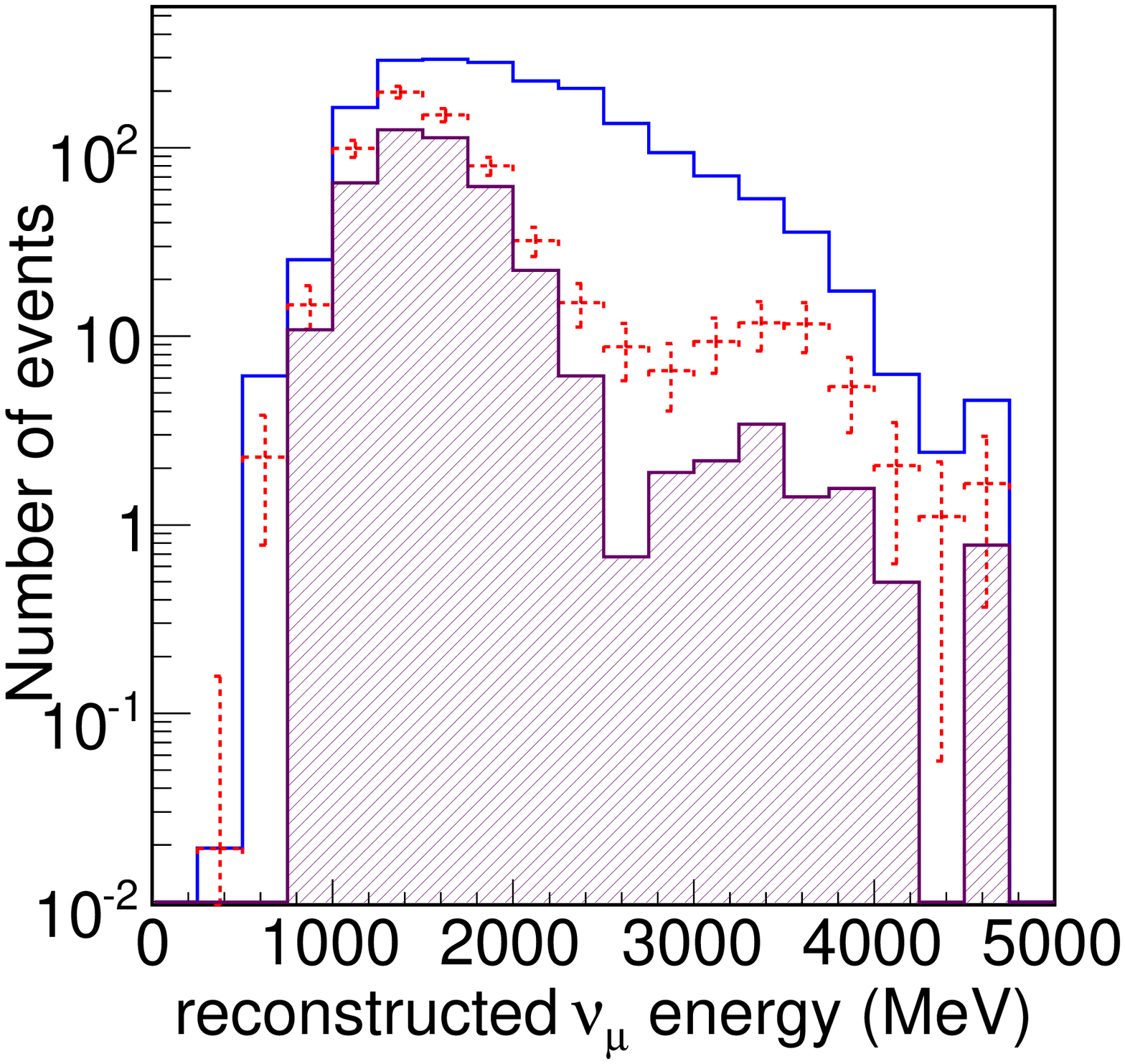,width=4.0in}
  \caption{Kinematically reconstructed $\nu_\mu$ spectrum for the LBNE
    beam. The full line shows the expectation assuming no oscillation
    while the dashed points with error bars show the typical expected
    spectrum with neutrino oscillations, with error bars corresponding
    to expected statistics. The hatched histogram shows the
    contribution from true CCQE events.\label{fig:pxspec}}
}

Figure~\ref{fig:pxspecproba} shows the ratio of the oscillated
Monte-Carlo to the non-oscillated Monte-Carlo. The error bars reflect
the amount of statistics available after 3 years. The oscillatory
shape is clearly visible, and shows the good energy resolution reached
with this technique. 
 Several sources of systematic errors are expected to affect the oscillatory
shape. In~\cite{proton}, a conservative 10\% error was used in the relative
CCQE selection efficiency. This also includes our imperfect knowledge of
the neutrino-nucleus cross-section. This error would distort the oscillatory
shape because non-CCQE background will have incorrectly reconstructed energy.
Another important source of error comes from inaccuracies and biases in
our proton track reconstruction. In \cite{proton} they were estimated to be 10\%,
which is also an overestimate.

\FIGURE{
  \epsfig{file=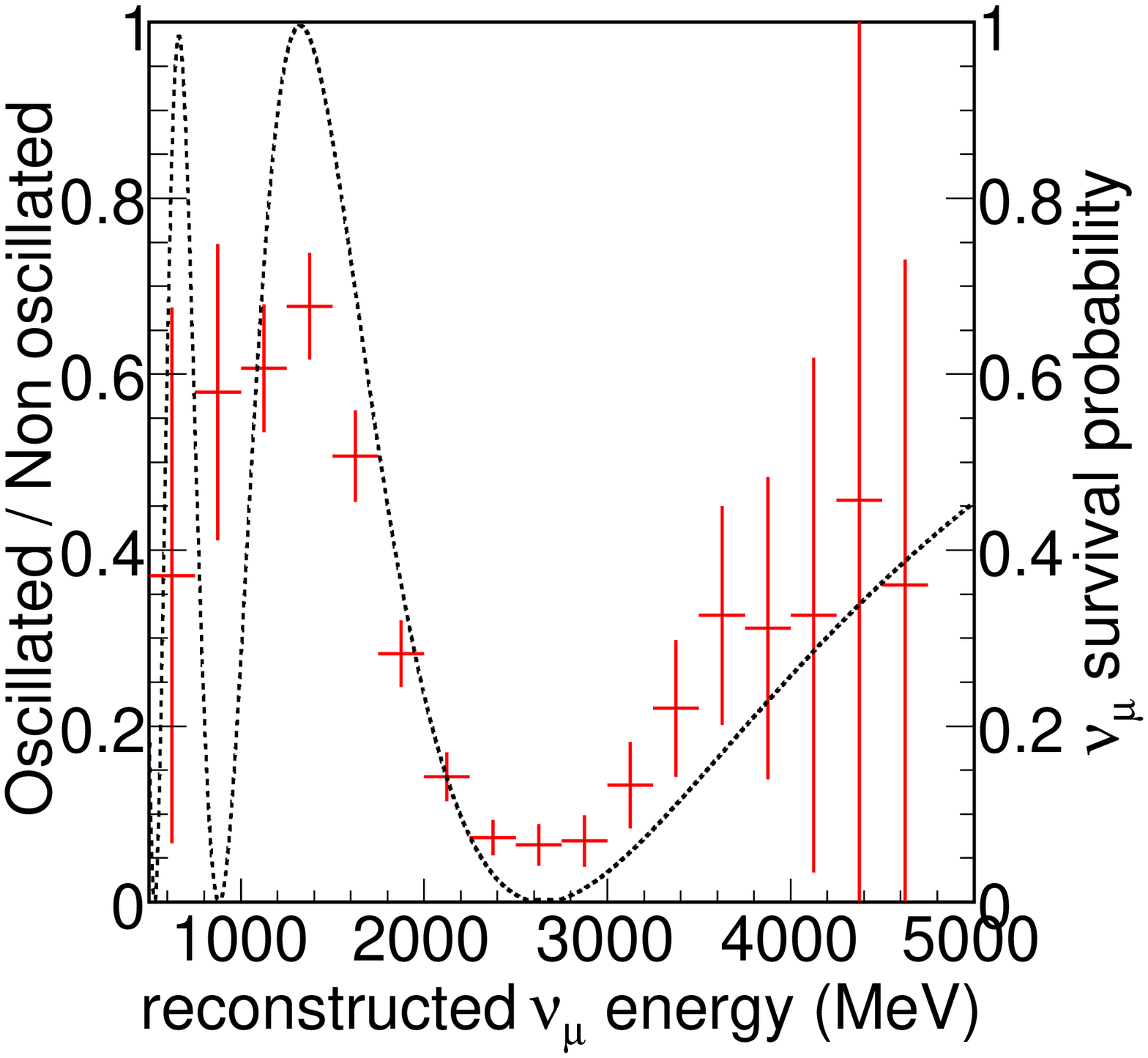,width=4.0in}
  \caption{Example of a measurement of the $\nu_\mu$ oscillation
    probability as a function of reconstructed neutrino energy: ratio of
    ``measured'' spectrum to expected spectrum. The overlaid dashed line
    is the result of an exact probability calculation, as a function of
    \emph{true} neutrino energy (explaining the disagreement).\label{fig:pxspecproba}}
}

\subsubsection{Summary of expectations in the LBNE project}

In summary, for this kind of wide-band beam and large detector, the tagged CCQE sample
has fairly large statistics of approximately 750-800 events, with a
CCQE purity of $\approx 60\%$.  The average resolution on the measured
neutrino energy for the sample is $\approx 15\%$.  
The number of tagged-CCQE events selected in the e-like sample will vary between
roughly 85 and 100 if $\theta_{13}$ varies from 0 to 4 degrees.
Although this would not by itself tightly constrain or measure
$\theta_{13}$, combining these fully reconstructed events with other
samples in a complete analysis would likely be quite helpful.

The $\nu$ to $\nu+\bar{\nu}$ ratio in the tagged CCQE sample
is estimated to be greater than 98.5\%, somewhat improving from the original
beam's content (95.5\%). Additionally, this technique could be used
with the anti-neutrino beam to tag and confirm the neutrino contamination
which should be approximately 30\%~\cite{bishai}. 

\subsection{Beta-beams}
\label{sec:beta}

A beta-beam or $\beta$-beam \cite{Zucchelli:2002sa} is a beam of \nue
or \nuebar obtained by accelerating radioactive ions (e.g.~$^{18}$Ne
or $^6$He) and letting them undergo $\beta$ decay in a storage ring
with long straight sections. The energy of the neutrino follows a
boosted $\beta$ decay spectrum, and the beam is pure \nue or pure
\nuebar depending on which element is used.

The performances depend on the end-point $E_0$ of the $\beta$-decay
spectrum, i.e.~on the ion; they also depend on the baseline $L$
between the accelerator and the detector, and the relativistic
$\gamma$ factor of the ions. Many combinations of these three
parameters have been considered for future facilities, and the reader
should consult e.g.~\cite{issphysics} for a summary of all the available
options.

In order to reach neutrino energies of a few GeV, so as to make
visible protons, the relevant options would be $L=700$ km and
$\gamma=350$, with $^{18}$Ne (for \nue production, with $E_0=3423.7$
keV) and $^{6}$He (for \nuebar production, with $E_0=3506.7$
keV). Such options (so called high-energy $\beta$-beams) could
correspond to a refurbished SPS (at CERN) or Tevatron (at FNAL). We
have assumed that the ion fluxes were $2.9\times 10^{18}$ decays per
year for He ions and $1.1\times 10^{18}$ decays per year for Ne ions,
following the EURISOL $\beta$-beam group quoted in
\cite{issphysics}. 
This configuration yields a flux peak at
$\approx$~1.6 GeV and a spectrum endpoint at $\approx$~2.4 GeV.
The total integrated flux per year is $6.5\times 10^{11}\ \mathrm{m}^{-2}$
for $\bar{\nu_e}$ and $2.4\times 10^{11}\ \mathrm{m}^{-2}$ for $\nu_e$.

\subsubsection{Event rates and purity of CCQE tagged events}

Re-weighting the SK atmospheric Monte-Carlo to the $\beta$-beam
spectra, and applying the same analysis as for the LBNE project we
have obtained the numbers in table~\ref{tab:bbeam} which correspond to
an exposure of 900 kton-year. The event rates are about four times 
higher than for the LBNE beam studied in the previous section 
due to the higher flux. Running in pure \nuebar mode would
ensure that there would be no neutrino CCQE events, which could
provide a separate measurement of the various background events that
contaminate the neutrino CCQE sample when running with $\nu_e$. Our
Monte-Carlo simulations show that $\approx 33\%$ of the tagged-CCQE
$\bar{\nu}$ events are $\bar{\nu}$-CCQE events, i.e.~$\bar{\nu}+p
\rightarrow \mathrm{lepton}+n$.  These events are tagged because the
outgoing neutron interacted hadronically with the detector's water and
produced a proton. Approximately $43\%$ of the $\bar{\nu}$ tagged-CCQE
sample comes from charged-current single-pion production (especially
$\bar{\nu}+p\rightarrow \mathrm{lepton}+p+\pi^-$, detecting the
outgoing proton), $\approx 13\%$ comes from neutral-current
single-pion production, and $\approx 9\%$ from charged-current
multi-pion production.

\TABLE{
  \begin{tabular}{lcccc}
    \hline\hline lepton flavor & ~Selected~ & ~true-CCQE~
    & ~Selected~ & ~true-CCQE~ \\ 
    photo-coverage & 40\% & 40\% &  20\% & 20\% \\ 
    \hline \multicolumn{5}{c}{$^{18}$Ne \nue beam}\\ 
    \hline
    e-like & $2954\pm 117$ & $2009\pm 95$ & $2301\pm 113$ & $1560 \pm 82$ \\ 
    $\mu$-like & $88\pm 20$ & $35\pm 12$ & $115\pm 23$ & $40\pm 13$ \\ 
    \hline\multicolumn{5}{c}{$^{6}$He \nuebar beam}\\ 
    \hline
    e-like     & $511 \pm 92$ & 0 & $392 \pm 82$ & 0\\ 
    $\mu$-like & $93 \pm 40$ & 0 & $44 \pm  26$  & 0\\ 
    \hline\hline
  \end{tabular}
  \caption{Expected event selection in a high-energy $\beta$-beam at
    $L=700$ km and $\gamma=350$ after the selection criteria defined in
    \cite{proton}. The columns labeled true-CCQE show the expected 
    numbers of true CCQE events in the tagged-CCQE sample.
    We have assumed standard 3-flavor neutrino
    oscillations, with $\theta_{13}=0$. }
 \label{tab:bbeam}
}

\subsubsection{Energy resolution of CCQE tagged events}

As with the wide-band superbeam described in the previous sub-section,
kinematic reconstruction of the incoming neutrino also improves the
energy resolution. Figure~\ref{fig:betaenresolusual} shows the energy
resolutions that can be expected with the $\nu_e$ $\beta$-beam,
selecting all single-ring e-like events, without any proton tagging,
which is the usual method for neutrino energy estimation in water
Cherenkov detectors. The neutrino energy is obtained from the beam
direction and the lepton information alone as in e.g. the K2K
experiment \cite{Ahn:2006zza}. This energy reconstruction method
assumes that all events are CCQE, and therefore any contamination with
non CCQE events leads to an error on the measured energy, usually an
underestimate. In figure~\ref{fig:betaenresolusual}, a narrow peak
corresponding to correctly identified CCQE events with good resolution
is visible, along with a large tail of non-CCQE events.  It can be
seen that two-ring events (dashed line) are almost always
mis-reconstructed with this method because they are largely
non-CCQE. They cannot be used when reconstructing the neutrino energy
with lepton information alone.

\FIGURE{
  \epsfig{file=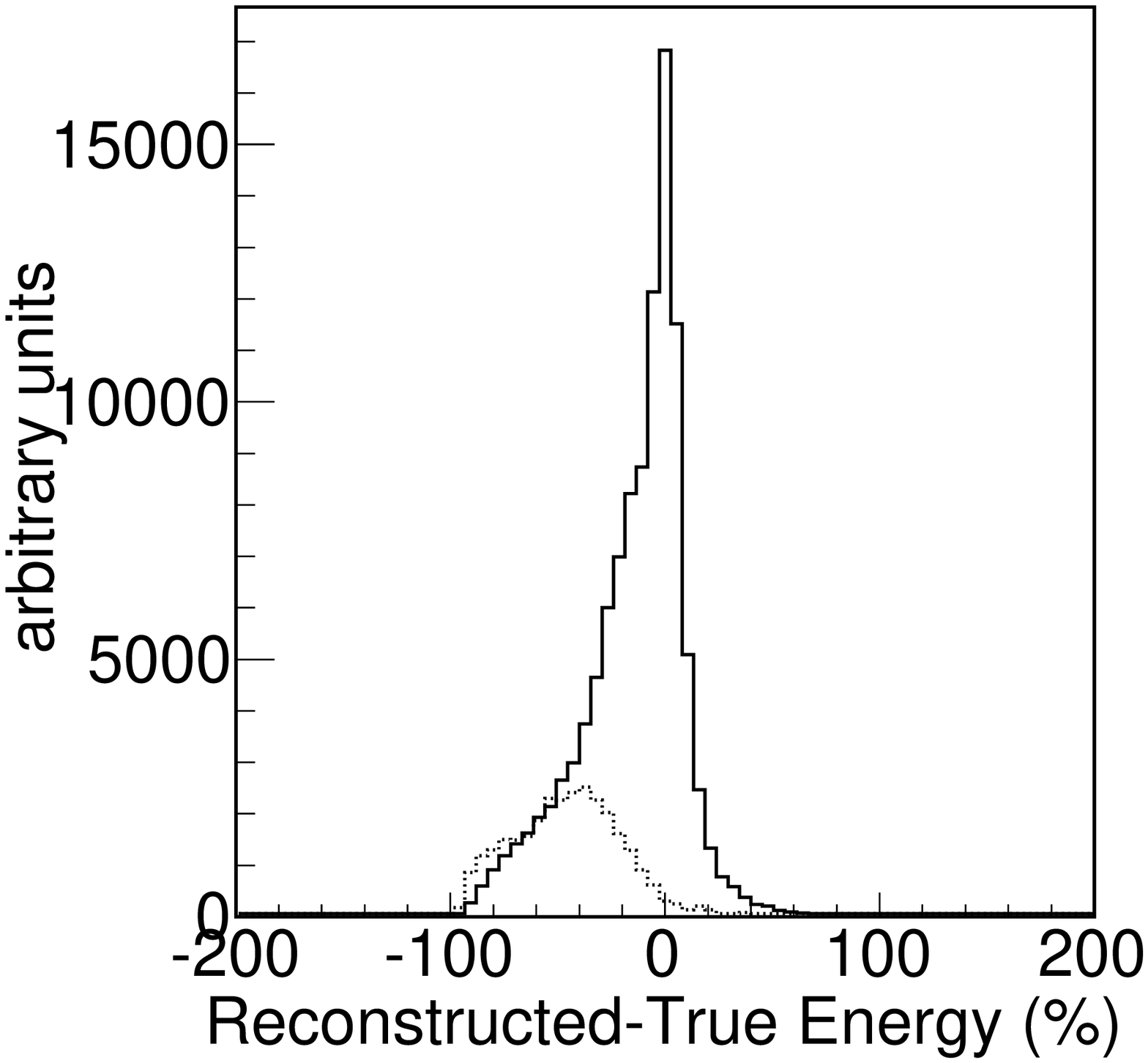,width=4in}
  \caption{Neutrino energy resolution for $\nu_e$ $\beta$-beam
    single-ring (full line) and two-ring (dashed line) e-like
    events. There is no proton tagging, and only the lepton
    information is used to estimate the neutrino energy.\label{fig:betaenresolusual}}
}

By contrast, figure~\ref{fig:betaenresol} shows the effect of using
proton tagging and full kinematic neutrino energy reconstruction with
the two tracks. The two methods for reconstructing the neutrino
energy, either full kinematic reconstruction with the proton and
lepton track, or reconstruction with the lepton track alone as shown
in figure~\ref{fig:betaenresolusual}, are compared.  In this figure
only tagged-CCQE events were used, in order to show the improvement
brought by proton tagging.  The energy resolution on the incoming
neutrino obtained using this technique as demonstrated in
figure~\ref{fig:betaenresol} is $\approx 11\%$.

\FIGURE{
  \epsfig{file=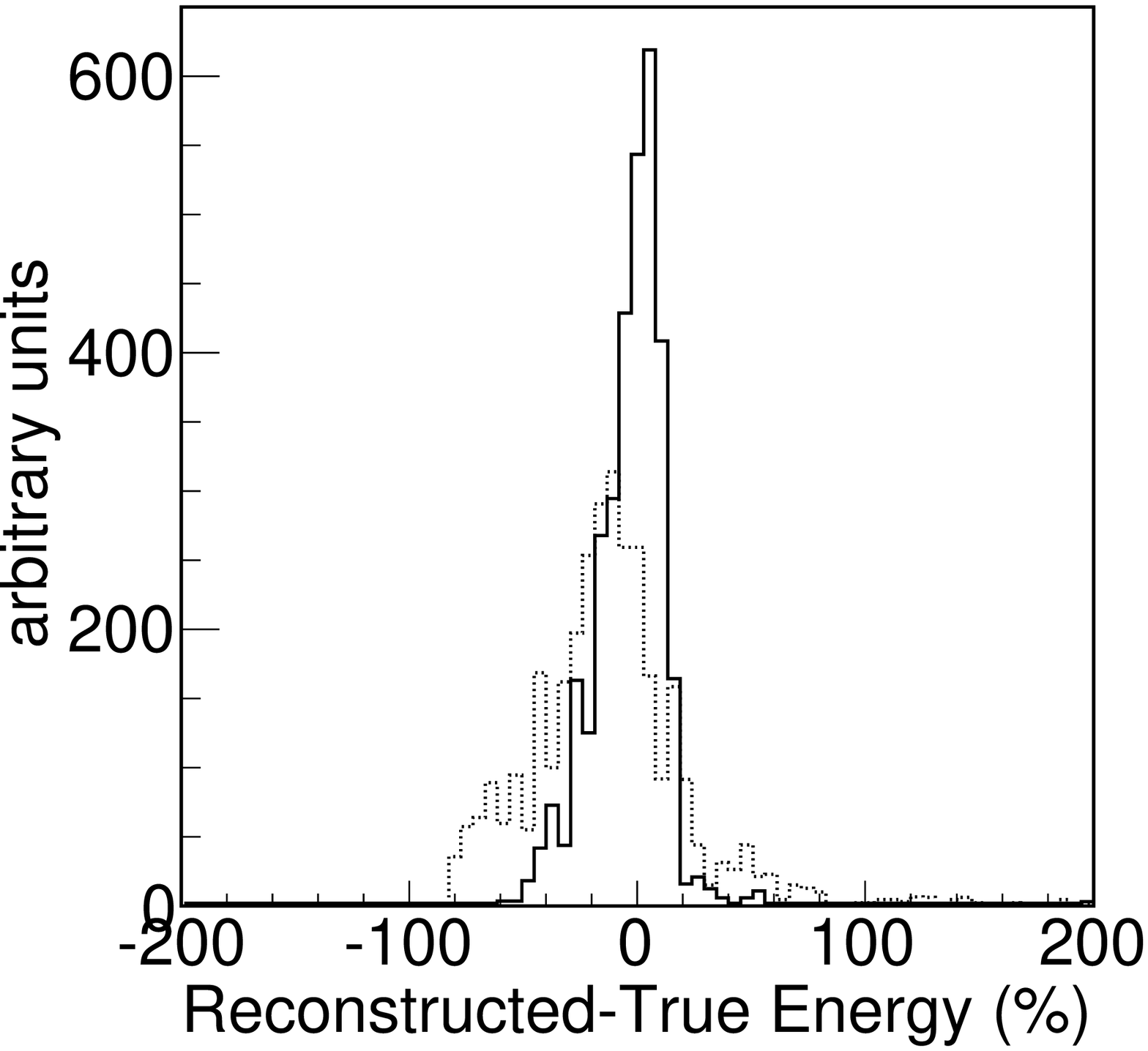,width=4in}
  \caption{Neutrino energy resolution for the tagged-CCQE sample
    from the $\nu_e$ $\beta$-beam.  The dotted line corresponds to
    reconstruction using the lepton information only, and the full
    line to full kinematic reconstruction using the proton and the
    lepton.\label{fig:betaenresol}}
}

\subsubsection{Summary of expectations in the beta-beam}

In summary, reconstructing the proton track eliminates non-CCQE events
very efficiently, makes use of the two-ring sample, and has good
energy resolution, but the requirements on proton visibility strongly
reduce the statistics.  The main interest of this technique for a
$\beta$-beam is to select a high resolution, $\approx 66\%$ pure CCQE
sample. 

\section{Conclusion}

In this paper, we studied the physics potential of a novel proton
identification technique in future large water Cherenkov detectors,
with atmospheric neutrinos and several possible neutrino beam
scenarios.  We used the SK-I (40\% photo-cathode coverage) and SK-II
(20\% photo-cathode coverage) simulations, which have been carefully
tuned on atmospheric neutrino data, and re-weighted them to match
various neutrino flux spectra.

For atmospheric neutrinos, a megaton scale Cherenkov detector
utilizing this technique would observe several hundred NC elastic and
CCQE events.  The energy resolution of the sample is excellent, but
the selected energy of the sample has the first oscillation maximum
near the horizon where the resolution is still poor. However, the
second oscillation maximum pattern might still be visible albeit with
relatively low statistics. There will be a quasi-pure neutrino sample, 
allowing further studies of mass hierarchy effects with
atmospheric neutrinos.  We find that, for this analysis, decreasing
the photo-cathode coverage from 40\% to 20\% decreases the event rate by
approximately 10\% for CCQE searches and 20\% for NC searches in
atmospheric neutrinos.  The purity of the CCQE sample is little
changed however.  Finally, sterile neutrino searches based on this
sample would be a valuable addition to more comprehensive searches.

We also studied how the proton identification and CCQE selection
techniques could be applied to neutrino beams. In order to have
significant statistics, the beam must have a high event rate in the
few GeV region, between 1 and 5 GeV. A wide band superbeam of the kind
envisioned in the LBNE project would produce a sample of a few hundred
events, while a beta-beam would produce a few thousand.  In both
cases, the newly selected sample provides a direct measurement of the
neutrino energy using event kinematics, with good resolution
(10-15\%).  In both cases the proton tag can be used to either purify
the neutrino beam or measure the neutrino contamination in the
anti-neutrino beam.

Water Cherenkov detectors are in general chosen because of their
cost-effectiveness, and because they are a mature technology. However
they are known to have several limitations (see
e.g.~\cite{issphysics}). Single ring event detection
is emphasized since, by using the lepton direction and incoming
neutrino direction alone, the energy of the neutrino can be
reconstructed. It is assumed that CCQE and non-CCQE events in this
sample cannot be separated. Consequently, the neutrino beams usually
studied with water Cherenkov detectors are often limited to
energies of 1 GeV or less to avoid contaminating the sample with too
many non-CCQE events which spoil neutrino energy reconstruction.

Moreover, water Cherenkov detectors are thought to be unable to detect
whether the incoming neutrino is a neutrino or an anti-neutrino, mainly
because this capability relies on measuring the sign of the outgoing
charged lepton and requires a magnetized detector.  These limitations
render them less attractive than other detection technologies
(e.g.~liquid argon) for CP violation studies, where these distinctions
can be crucial.

We have shown here that these limitations can be partially lifted by
careful analysis of the Cherenkov light patterns: the non-CCQE
contamination can be reduced, CCQE events can be tagged, and the
neutrino can be kinematically reconstructed, ensuring at the same time
that it is a neutrino and not an anti-neutrino.  We have also shown
that beam spectra with peak energies in the few GeV region are
needed for these studies.  The exact quantitative effect on the
sensitivity to the mass hierarchy or $\delta_{CP}$ is beyond the scope
of this work and depends on the exact beam and detector configuration
along with their locations. However, we have shown it is possible to
obtain a few hundred atmospheric neutrino events, and depending on the
beams, from a few hundred to a few thousand events in a megaton scale water
Cherenkov detector.  

As an example of another possible impact of this technique we note
that in \cite{Huber:2008yx}, the authors show in a low-energy neutrino
factory, $\nu$ to $\nu+\bar{\nu}$ ratios of $50-90\%$ are enough to
reach good sensitivity to CP-violation and the mass hierarchy,
provided that $\sin^2 2\theta_{13}>10^{-3}$.  Also,
in~\cite{Schwetz:2008tc}, the author explains that even a modest
statistical separation between neutrinos and anti-neutrinos in a large
detector with particle identification can have dramatic impact on the
sensitivity to the mass hierarchy either using atmospheric neutrinos
alone, or combined with long-baseline results.  We hope that
these new capabilities of water Cherenkov detectors will be useful
when designing these future facilities.

\section{Acknowledgments}

The authors would like to thank the Super-Kamiokande collaboration for
the use of simulation tools and Monte Carlo samples employed in this
work.  This paper, and all of the results and conclusions presented
here, are the sole work of the authors themselves, and not of the
Super-Kamiokande collaboration.

We gratefully acknowledge individual support by the United States
Department of Energy (grant DE-FG02-91ER40665-A) and the United States
National Science Foundation (grant 0349193). We would also like to
thank M.~Bishai and F.~Dufour for providing beam fluxes and
normalization factors for the Project-X based beam.  We would like to
thank C.~Ishihara, as well as P.~Huber and W.~Winter for useful
discussion about beta-beam fluxes.

\bibliographystyle{JHEP.bst}
\bibliography{bibliography2}

\end{document}